\documentclass[
showkeys,
twocolumn,
superscriptaddress,
showpacs,
amsmath,amssymb,
aps,
pre,
]{revtex4-1}
\usepackage{natbib}
\usepackage{graphicx}
\usepackage{dcolumn}
\usepackage{bm}

\begin{document}

\title{Absorbing phase transition in a unidirectionally coupled 
layered network}
\author{Manoj C. Warambhe}
\author{Ankosh D. Deshmukh}
\author{Prashant M. Gade}
\affiliation{Department of Physics,
Rashtrasant Tukadoji Maharaj Nagpur
University, Nagpur-440033, India.
 }
 
\date{\today}

\begin{abstract}

We study the contact process on layered networks in which each layer is unidirectionally coupled to the next layer.  Each layer has elements sitting on i) Erd{\"o}s-R{\'e}yni network, ii) a $d$-dimensional lattice. The layer at the top which is not connected to any layer. The top layer undergoes absorbing transition in the directed percolation class for the corresponding topology. The critical point for absorbing transition is the same for all layers. For Erdos-Reyni network order parameter $\rho(t)$ decays as  $t^{-\delta_l}$ at the critical point for $l'$th layer with $\delta_l \sim 2^{1-l}$. This can be explained with a hierarchy of differential equations in the mean-field approximation.  The dynamic exponent $z$ is $0.5$ for all layers and the value of $\nu_{\parallel}$ tends to 2 for larger $l$. For a d-dimensional lattice, we observe stretched exponential decay of order parameter for all but top layer at the critical point.

\end{abstract}


\maketitle
\begin{center}
{\textbf{I.  Introduction}}
\end{center}
Identification of underlying topological
structure for complex systems\cite{nak03} has led to the new 
branch of $\lq$network
science'\cite{bor07}. Several researchers have studied different 
properties of real-life networks and proposed models.
Most popular among these models
are scale-free\cite {hei06} and small-world networks\cite{new11}.
The studies on networks  helped to a better understanding
for phenomena as diverse as
the spreading of diseases in the population, information
processing in gene circuits and biological pathways.
It has also helped in understanding transport properties on several
man-made system.

Another model 
which has attracted attention recently has been
multiplex network.
It models multiple levels of interaction in a given network.
One example is a social media network\cite{zha18,kan15} where individuals
 are connected by twitter, facebook, whatsapp, etc. The same individual could be connected to different individuals in various layers and there is certain information flow in the layers.
Another example is traffic network\cite{tia16} where people travel using
various modes of travel such as tram, bus, etc.  
In spread of diseases\cite{san14,de17},  empirical studies on different
strains of disease or different diseases have shown 
the necessity of modeling
the underlying network as a multiplex network.
In a multiplex network, the interaction between the nodes
is described by a single layer network and the different layers of networks
describe the different modes of interaction.
Various properties
such as properties of random walk\cite{guo16} on these networks,
eigenvalue\cite{gom13} and eigenvector structure of these networks, spread
of infection on such networks etc. have been investigated.

In this work, we study a simplified model of multilayer networks where
all layers have the same type of connectivity within a given layer.
Every agent is connected to the agent in the next layer in a unidirectional
manner. We study the contact process on this network. For
low infection probability $p$, the infection
dies down and number of infected individuals goes to zero. For higher $p$, the fraction of infected individuals
tends to a constant. Usually, this is an absorbing transition
in the universality class of directed percolation. We study this model on
the network mentioned above and find that the
nature of decay of order parameter at the critical
point changes from layer to layer. Interestingly, for a random network,
we observe a power-law decay of order parameter with different exponents for
different layers. On the other hand, for 1-d or 2-d basic networks, we find
that the decay is well described by stretched exponential at the critical point for all but top layer.\\

\begin{center}
{\textbf{II.  The Model}}
\end{center}
First, we consider a multiplex network with $L$ layers each having $N$ agents.
Each layer has
Erd{\"o}s-R{\'e}yni  type random network, {\it{i.e.}}
each site is coupled to $k$ randomly chosen sites in the same layer for
top layer and same connectivity is repeated for all $L$ layers.
Each site is 
connected to the previous layer unidirectionally. 
Each $m^{th}$ site in $j^{th}$ layer is connected
to  $m^{th}$ site in $j-1^{th}$ layer of the lattices in
unidirectional way for $j>1$. The top layer
($j=1$) is not connected to layer.
The representative picture of random network topology  for
only two layers and for $k=2$ is shown in Fig.1 (a).
Apart from a random network, we have also considered cartesian
lattice as a network for the top layer in later sections.
Representative multiplex structure for 1-D network for 4 layers
is shown in Fig.1 (b).
\begin{figure}[hbt!]
\scalebox{0.19}{
        \includegraphics{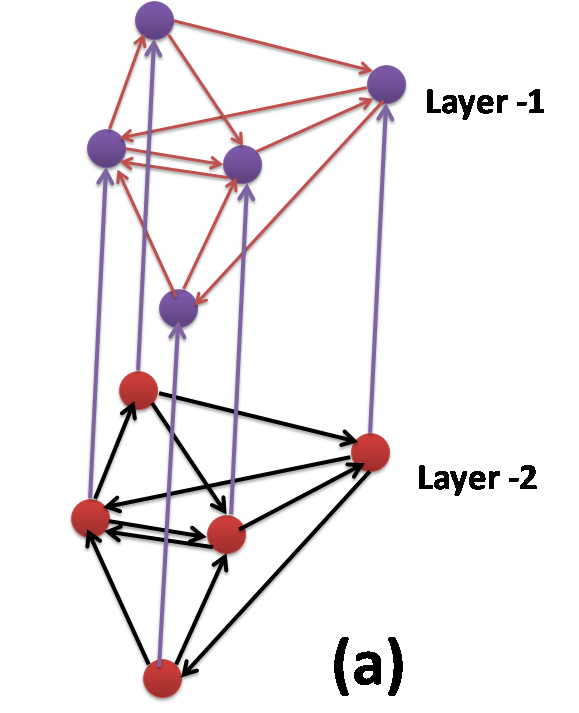}
}
\scalebox{0.34}{
        \includegraphics{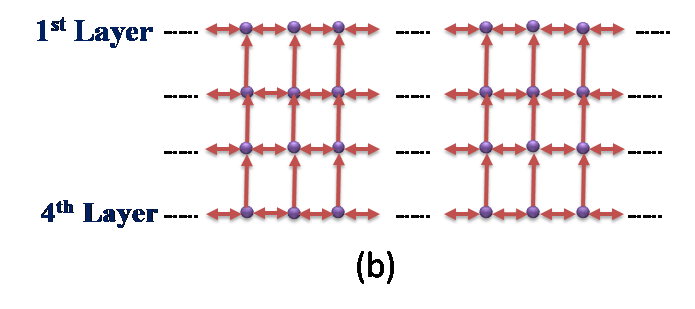}
}
\caption{(a)Topological representation of random network system.
 (b)Topological representation of 1-D network system.}
\label{Fig:1}
\end{figure}
We have carried out extensive numerical simulations for
contact process on
above random multiplex network where the top layer is a random network with $k=4$.
We define the contact process on this network as follows. We associate
variable $x^j_m(t)$ to $m$'th
site on $j$'th layer of this $NL$ dimensional multiplex where
$L$ is a number of layers each of which has $N$ sites. 
Initially, we assign  $x^j_m(0)=0$ or
$x^j_m(0)=1$ with equal probability.
We define $s^j_m(t)$ as sum of
$x^j_m(t)$ which are connected to $x_m^j$.
 The evolution proceeds in a
synchronous manner as $x^j_m(t+1)=1$ with probability $p$ if
$s^j_m(t) \ne 0$ and 0 otherwise. In other words, each site
becomes active with probability $p$ if any of the sites it is
connected with is active.  Being a contact process,
this model shows the transition to an absorbing state. If all sites in the
multiplex become inactive, they remain so forever.
Furthermore, we observe another feature.
Due to unidirectional connection between layers,
if the entire  top layer become inactive,
it remains so forever because it is not
connected to any other layer.
Similarly, if all sites in the top two layers become inactive, 
they stay inactive
regardless of the presence of active sites in the next layers.
On the other hand, $j$th inactive layer can become active if
there are active sites in any $l$th layer such that $l<j$.
We expect the value of $p_c$ to be the same for
the entire lattice as it is for the top layer.
The reason is simple. Below $p_c$, the top layer will become inactive. Now
immediate next layer is the top layer for all practical
purposes and will become inactive and so on.

For a random network with $k$ neighbors, we expect the absorbing
state for $kp<1$ in the mean-field limit. Thus we estimate $p_c=1/k$.
For $k=4$, we numerically obtain $p_c=0.25000\pm 0.00015$ which is
close to this approximation. It is expected that
the dynamic phase transition for random connectivity will be in the same
universality class as the mean-field class.
For non-equilibrium phase transitions, this expectation is not always
fulfilled\cite{sudeshna,ashwini}.

The top layer is not connected to any layer and
thus the critical point as well as
the critical exponents for  absorbing phase transition in the top layer
must be in the same universality class as the absorbing
phase transition for that connectivity. As mentioned above,
this is also the critical point for the entire multiplex structure.
However, we may question how the critical exponents (if any)
change for layers below the top layer.\\

\begin{center}
{\textbf{A.  Erd{\"o}s-R{\'e}yni network}}
\end{center}
We study the 6-layers random network in which we study
the absorbing phase transition using order parameter $O_l(t)$ which
is a fraction of active sites in $l^{th}$  layer as a quantifier.
We indeed observe power-law decay of order parameter at the critical point
$p=p_c$ for all $l$. The order parameter
goes like $1/t^{\delta_l}$ for each layer $p=p_c$.
The power-law exponent value for the top layer is close to
$\delta_1=1$.  For layer below the top layer is $\delta_2=0.5$ and so on.
The magnitude of the power-law exponent of a layer decreases as we go down
the layers.  The value of $\delta_l$ for $l^{th}$ layer is half
the value for $(l-1)^{th}$ layer.  Due to continuous infusion of infection
from layers above, the inactivation rate becomes slower for larger $l$.
This is shown in figure 2(a). An excellent power-law is obtained
with $\delta_l=2^{1-l}$ for $l>1$ while for $l=1$, we
obtain value $\delta_1=1$ which is equal to expected
mean-field value 1.  This behavior is confirmed
by plotting $O_l(t)t^{\delta_l}$ as a function of
time $t$ and independent fits (see fig. 2(b)). These values are confirmed
within $1\%$. We note that $O_l(t)t^{\delta_l}$
is constant in time over a few decades.  
While the exponent in the top-layer is an expected exponent in
mean-field class, other exponents are new.
\begin{figure}[hbt!]
\scalebox{0.3}{
        \includegraphics{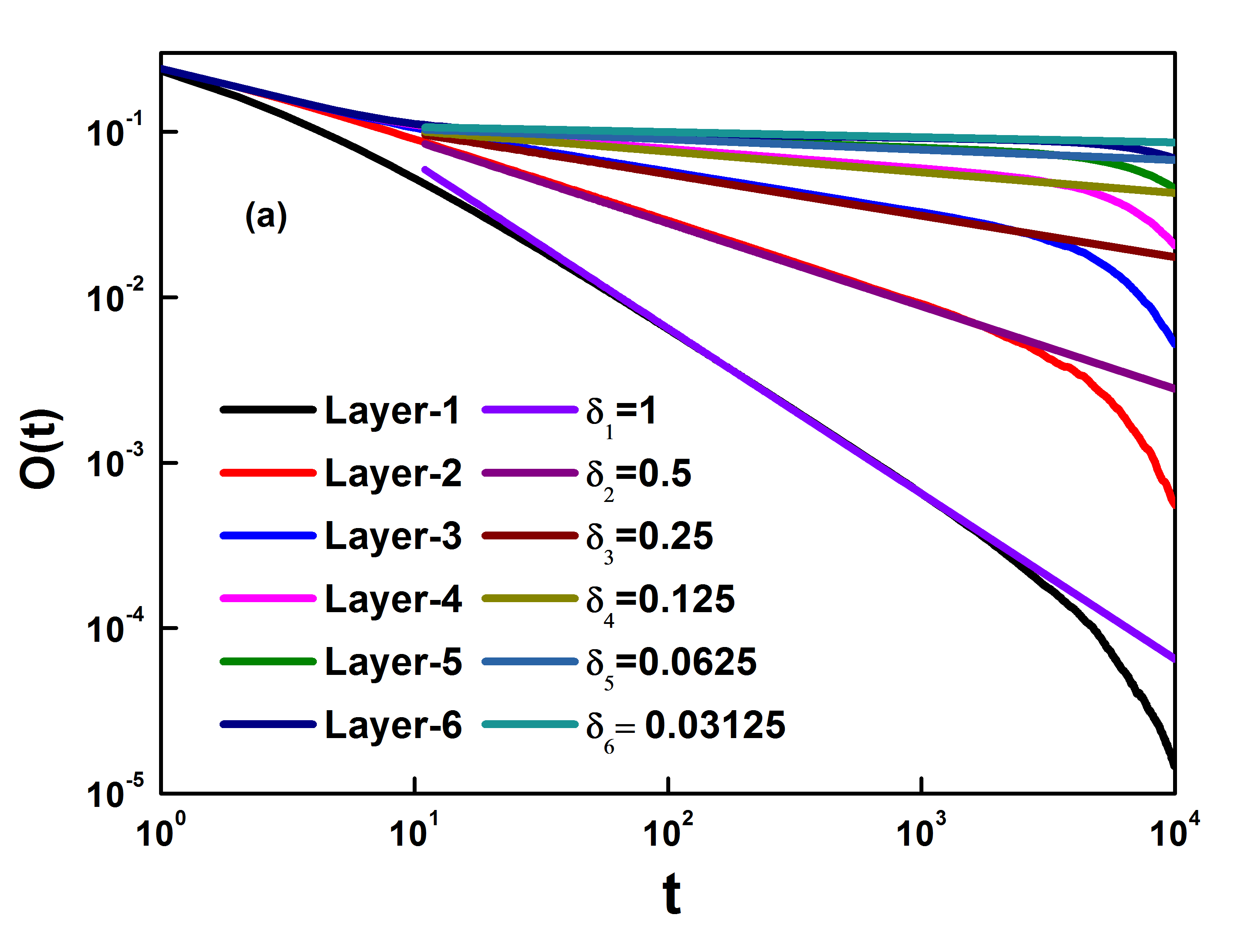}
}
\scalebox{0.3}{
        \includegraphics{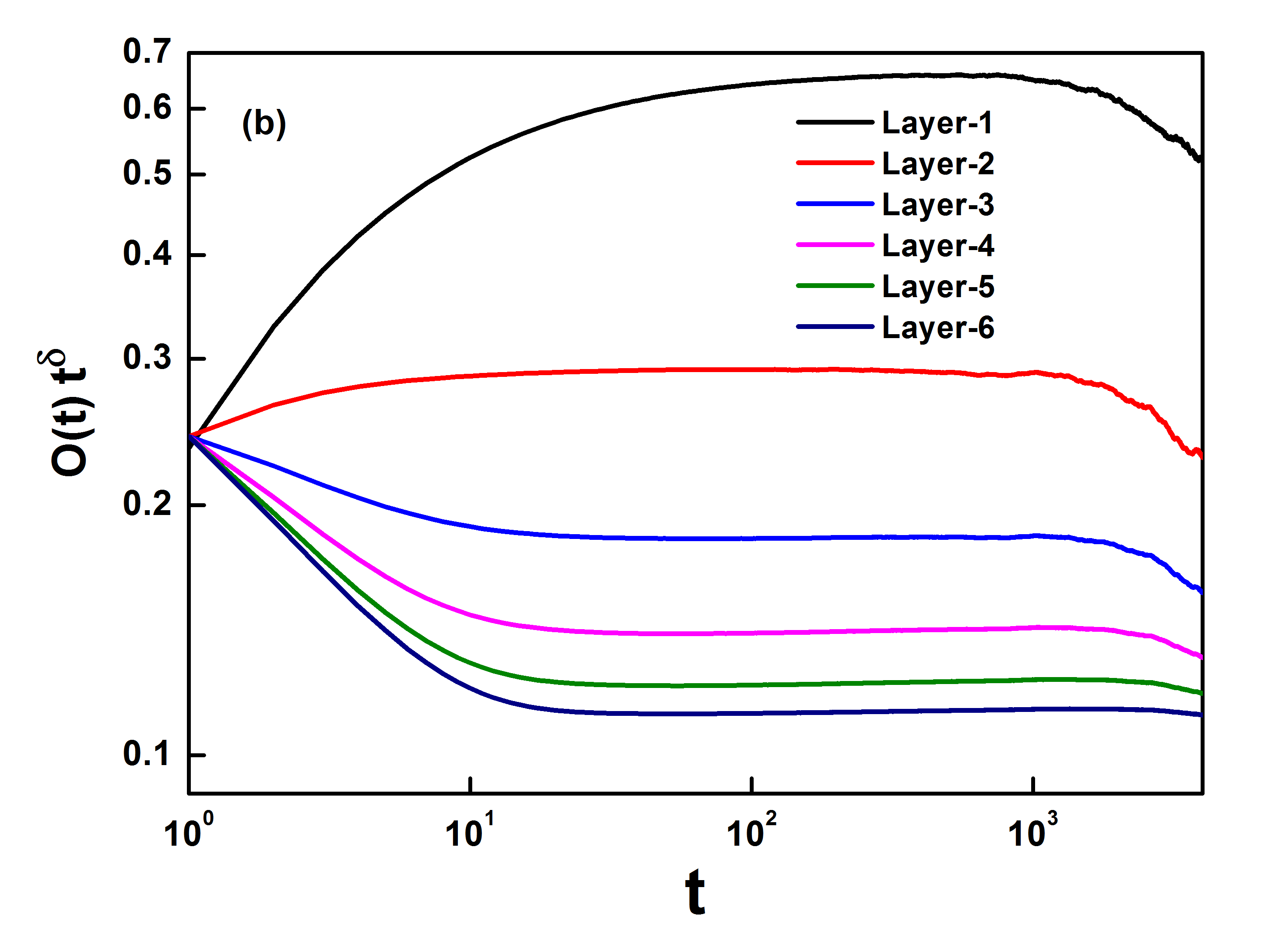}
}
\caption{a) We plot order prameter $O_l(t)$ as function of time $t$
for various layers (from bottom to top) of random network at $p=p_c=0.25$.
The lattice size is $N= 8 \times 10^{6}$.
The decay exponent is
given from $\delta_l = 1 - 2^{-l} $ and the fit is shown as guide to eye.
b) The quantity $O_l(t)t^{\delta_l}$ is plotted as a fuction of $t$ (from top to bottom).
We observe  that this quantity is a constant in time confirming $\delta_l$.}
\label{Fig:2}
\end{figure}
We study the finite-size scaling at the critical point for
different layers.
 We simulate for
$N=$ 100, 200, 400, 800, 1600, 3200, 6400, 12800, 25600, 51200. We have
 average over $2\times 10^6$ or more configurations
for $N\le 12800$ and over $2\times
10^5$ configurations for $N\ge 25600$.
We obtain finite-size scaling for every layer in the network.
The dynamical exponent value for all layers is the same and
has the value $z=0.5$.
The finite-size scaling for each layer is
shown in figure 3.
\begin{figure}[hbt!]
\scalebox{0.16}{
\includegraphics{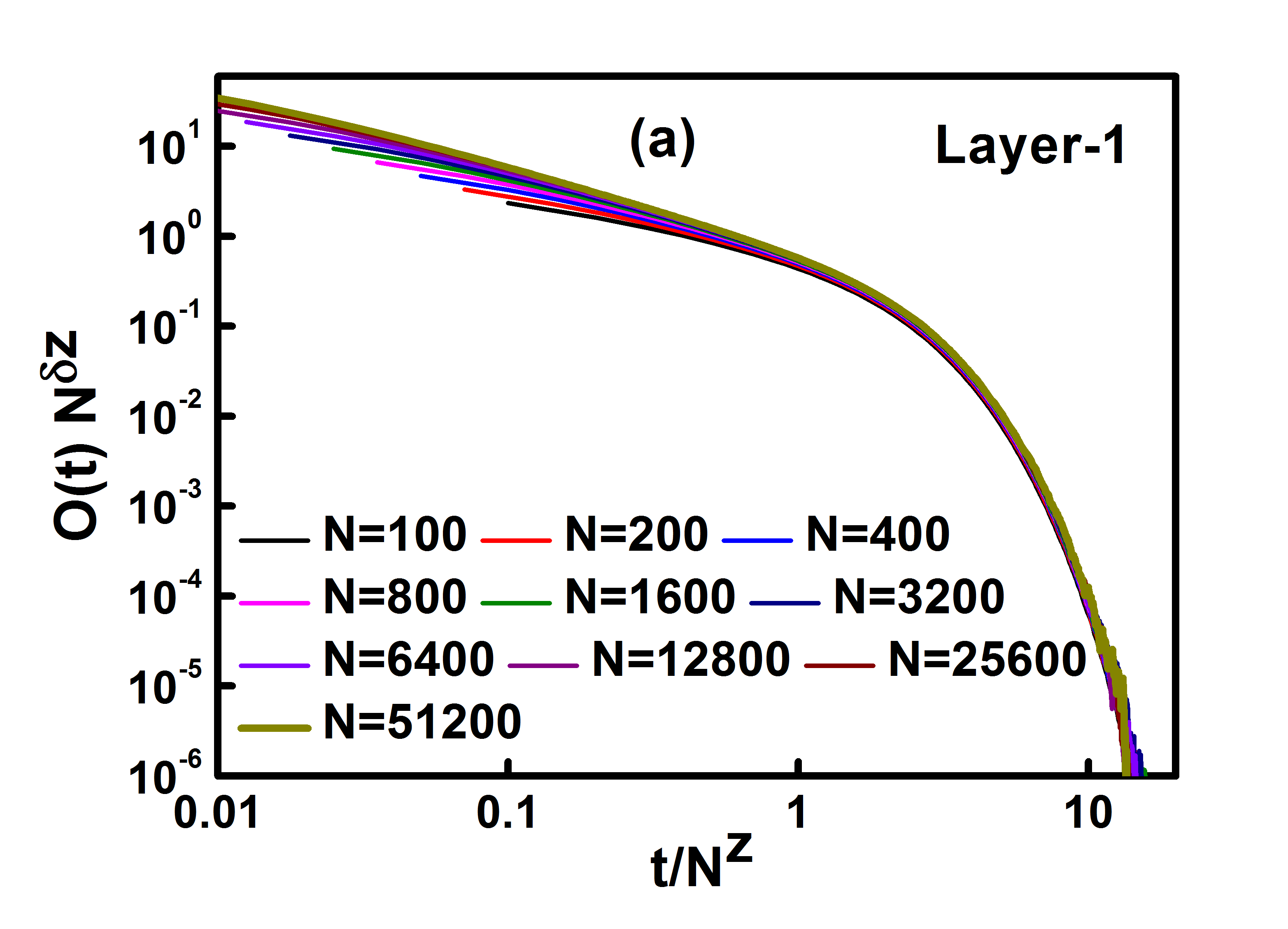}
}
\scalebox{0.16}{
\includegraphics{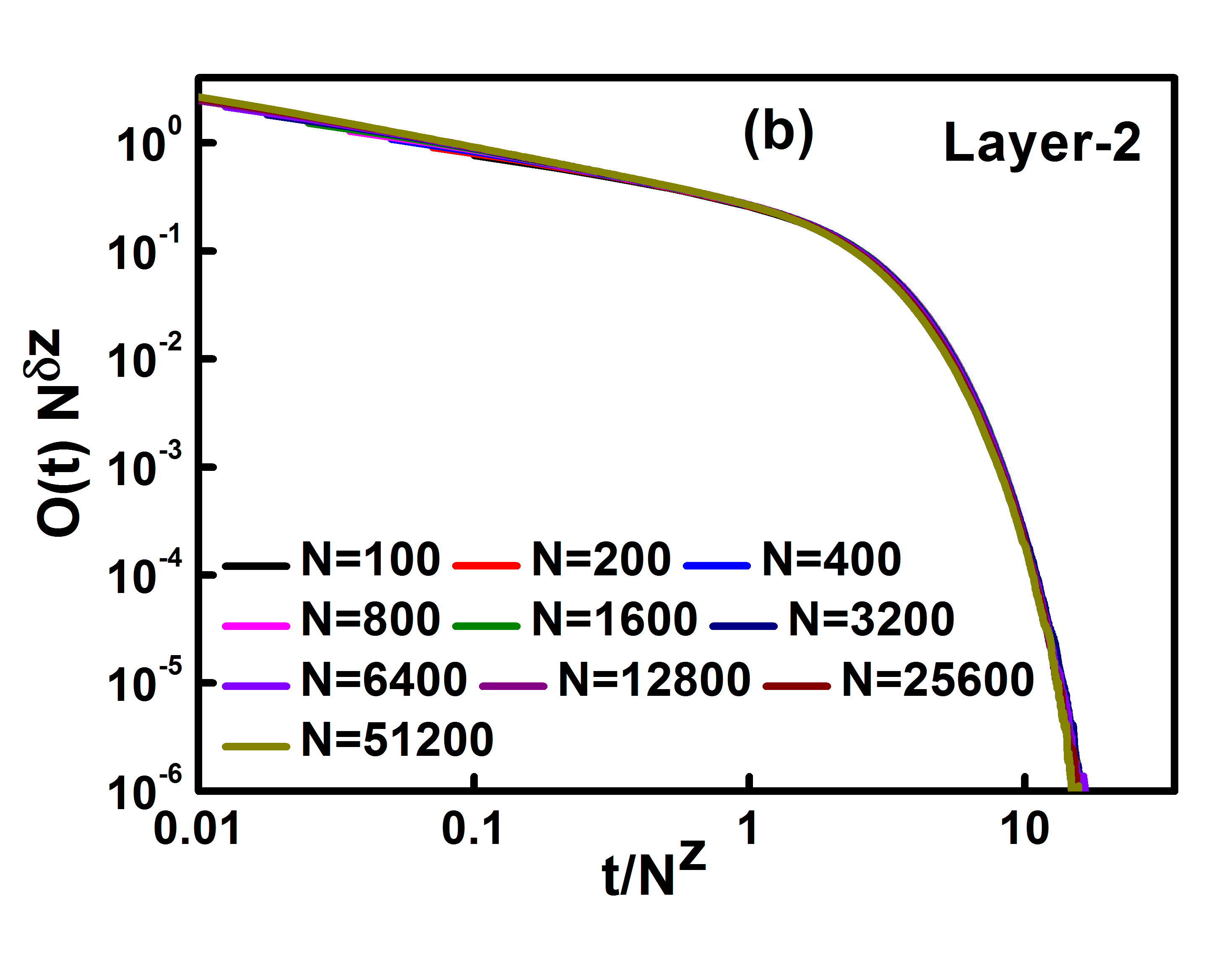}
}
\scalebox{0.165}{
\includegraphics{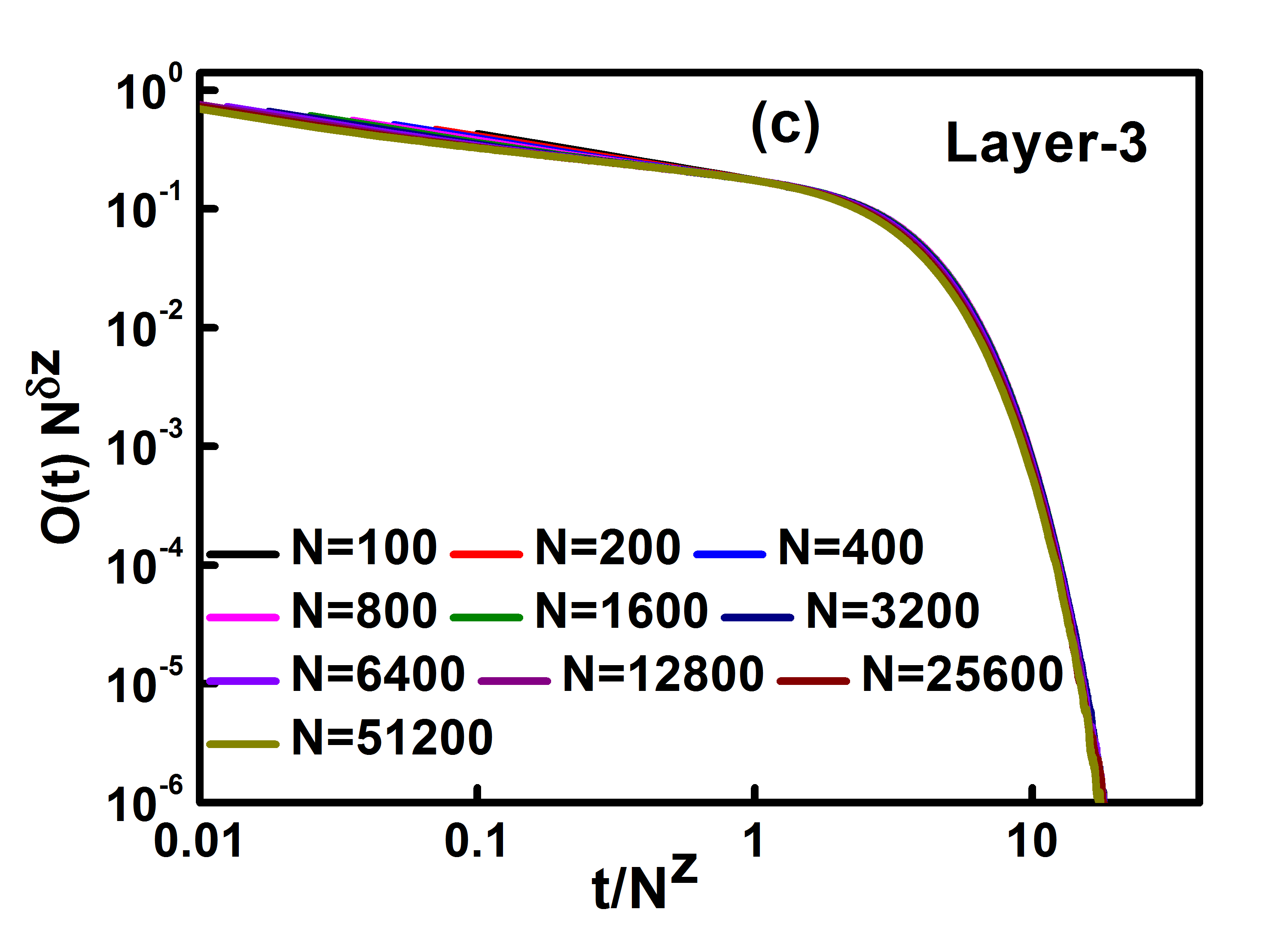}
}
\scalebox{0.16}{
\includegraphics{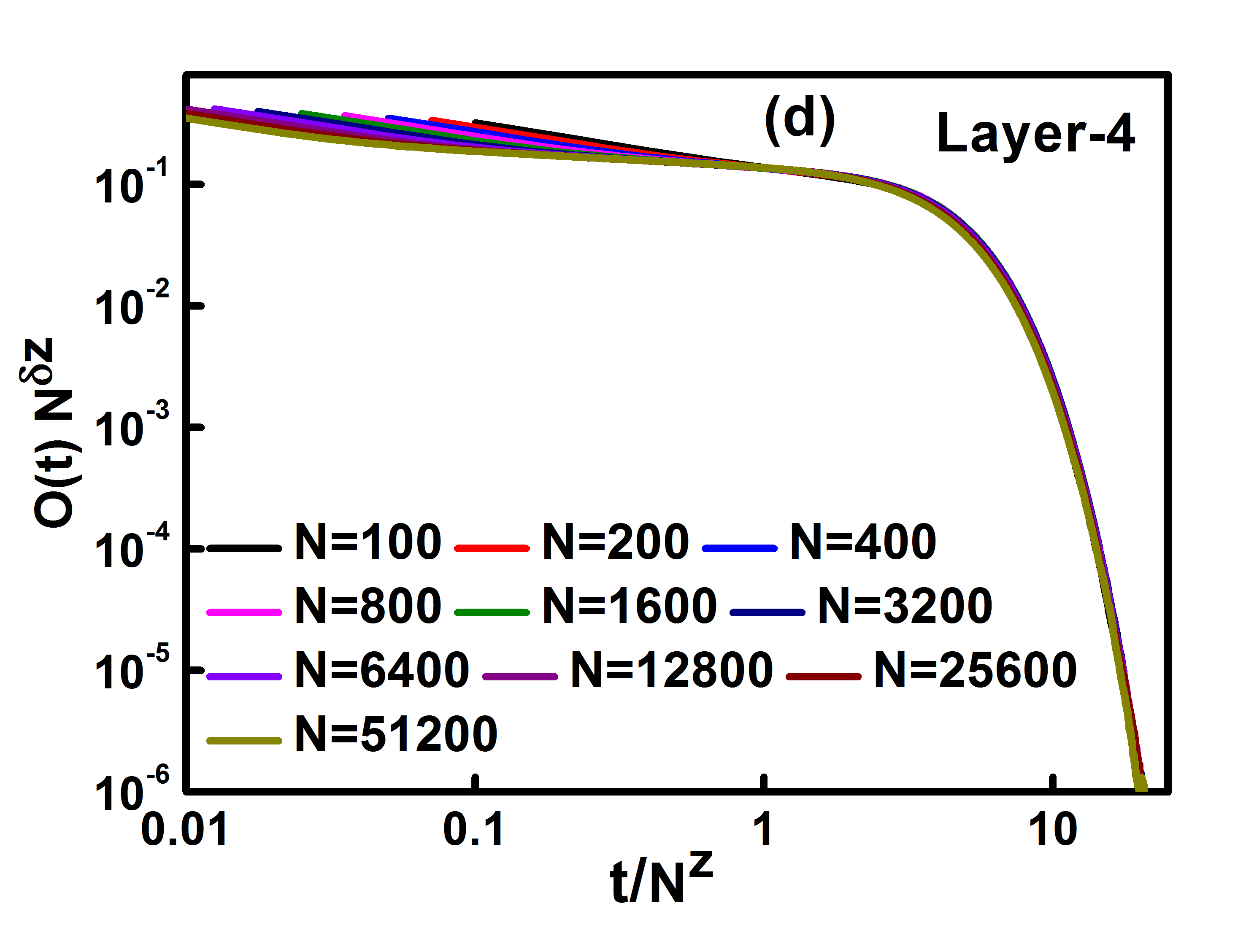}
}
\scalebox{0.16}{
\includegraphics{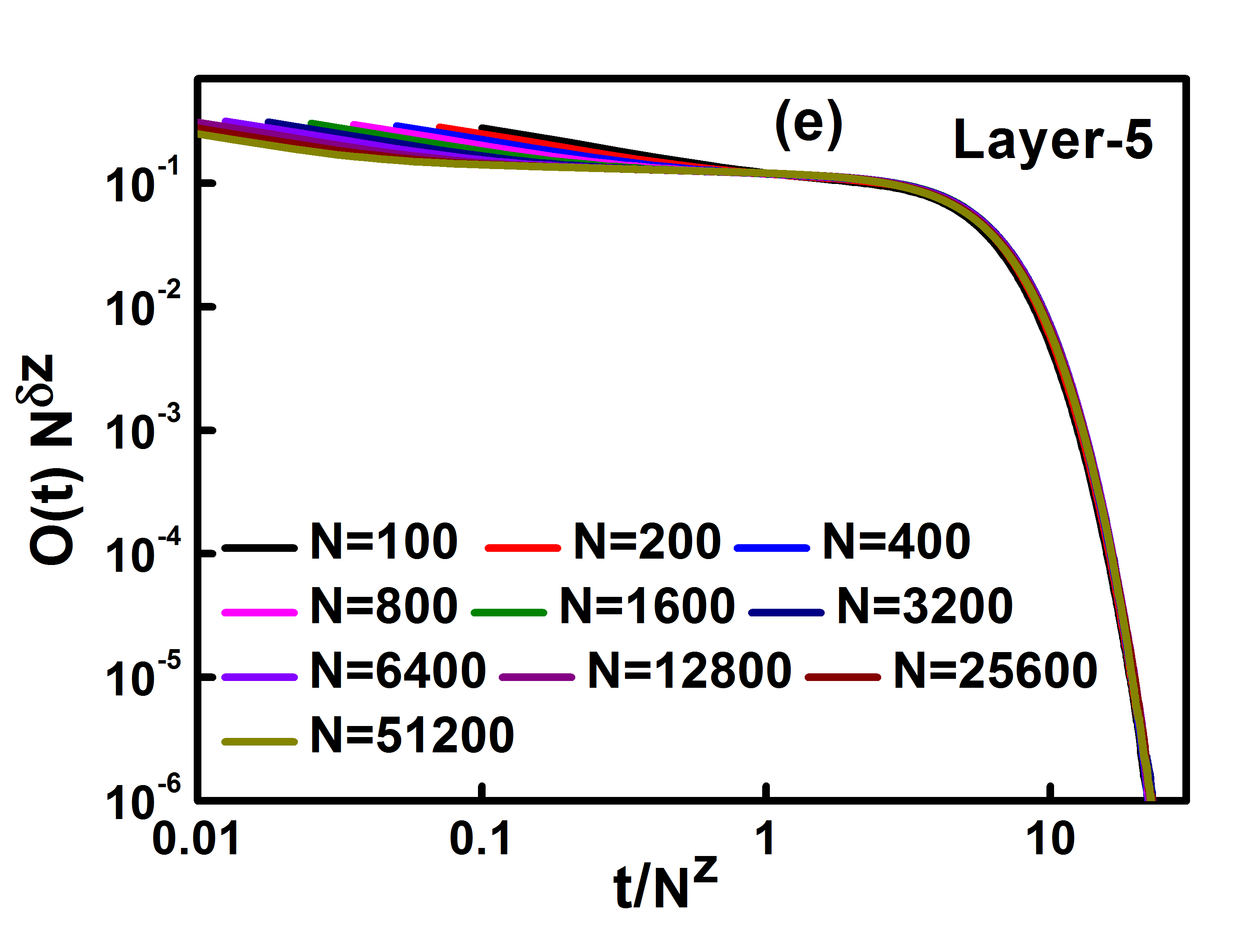}
}
\scalebox{0.16}{
\includegraphics{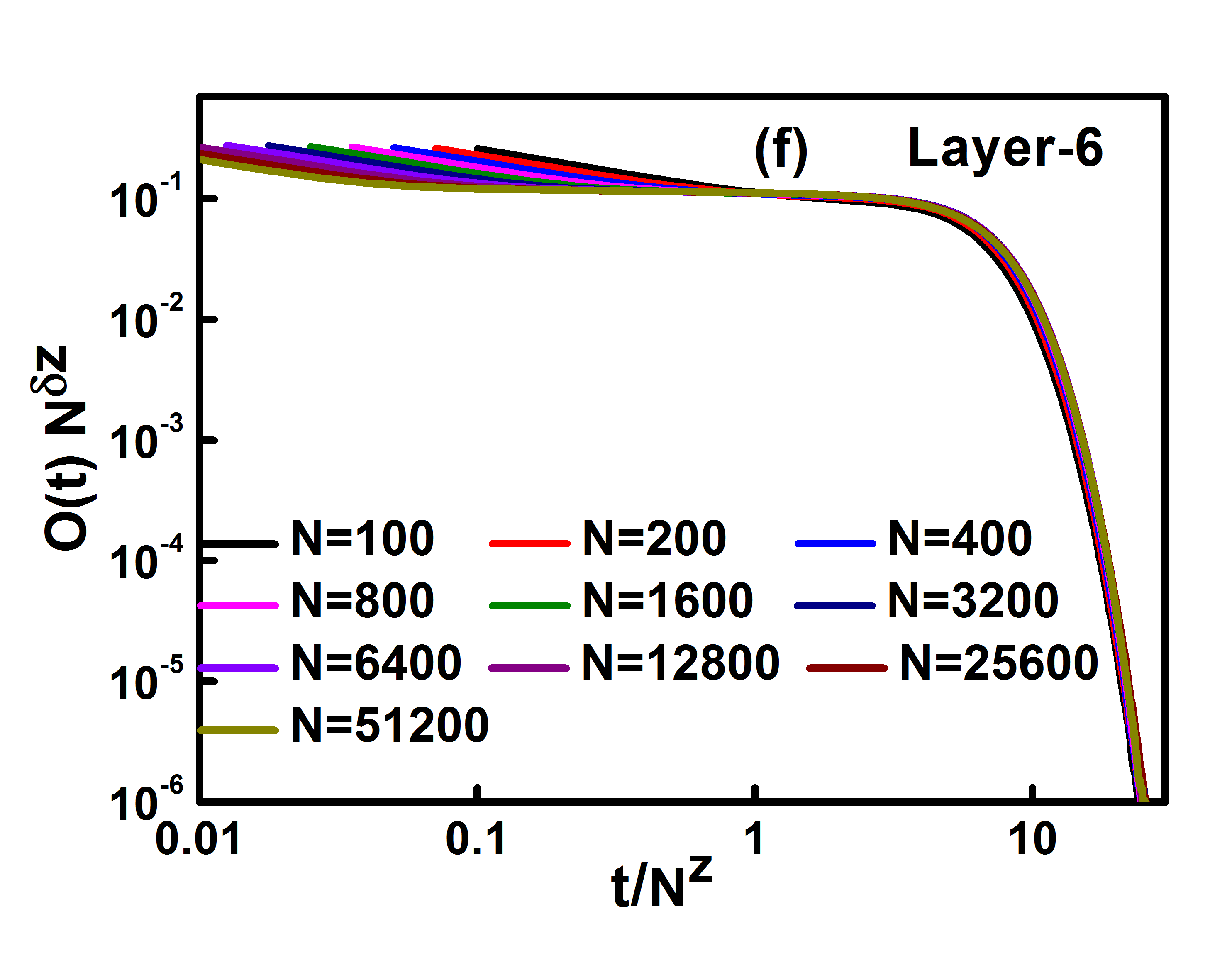}
}
\caption{For random network, we carry out  finite size scaling
by plotting plot
$O_j(t)N^{z\delta_l}$ 
as a function of $t/N^{z}$
for different system size $N$
at $p=p_c = 0.25$ where $\delta_l=2^{1-l}$ and $\delta_1=1$.
The value of dynamical exponent $z=0.5$ is same for all layers.
a) Layer-1 b) Layer-2 c) Layer-3 d) Layer-4 e) Layer-5 f) Layer-6}
\label{Fig:3}
\end{figure}

We expect the asymptotic value of order parameter to scale as
$O_l(\infty)\propto \Delta^{\beta_l}$  where $\Delta=\vert p-p_c \vert$
and $O_l(\infty)$ is the fraction of
active sites in $l^{th}$ layer.
We also note that $\beta_l=\nu_{\parallel,l}\delta_l$
We carry out simulations for
$N=8\times 10^6$ and average over more than $80$ configurations (see fig.4).
(We fit the function $O_l(\infty)\propto a \Delta^b$  using 
fit function in  gnuplot
 and values of $b$ obtained from fiting is closely match with 
the exponent $\beta_l$ obtained using visual fit.)
We find that $\nu_{\parallel,1}=\delta_1=1$ for
first layer which are mean-field values.
However for $l\ne 1$, $\nu_{\parallel,l}>1$ and $\beta_l\ne \delta_l$
for $l>1$. In fact $\nu_{\parallel,l}\rightarrow 2$ for higher layers.
\begin{figure}[hbt!]
\scalebox{0.3}{
\includegraphics{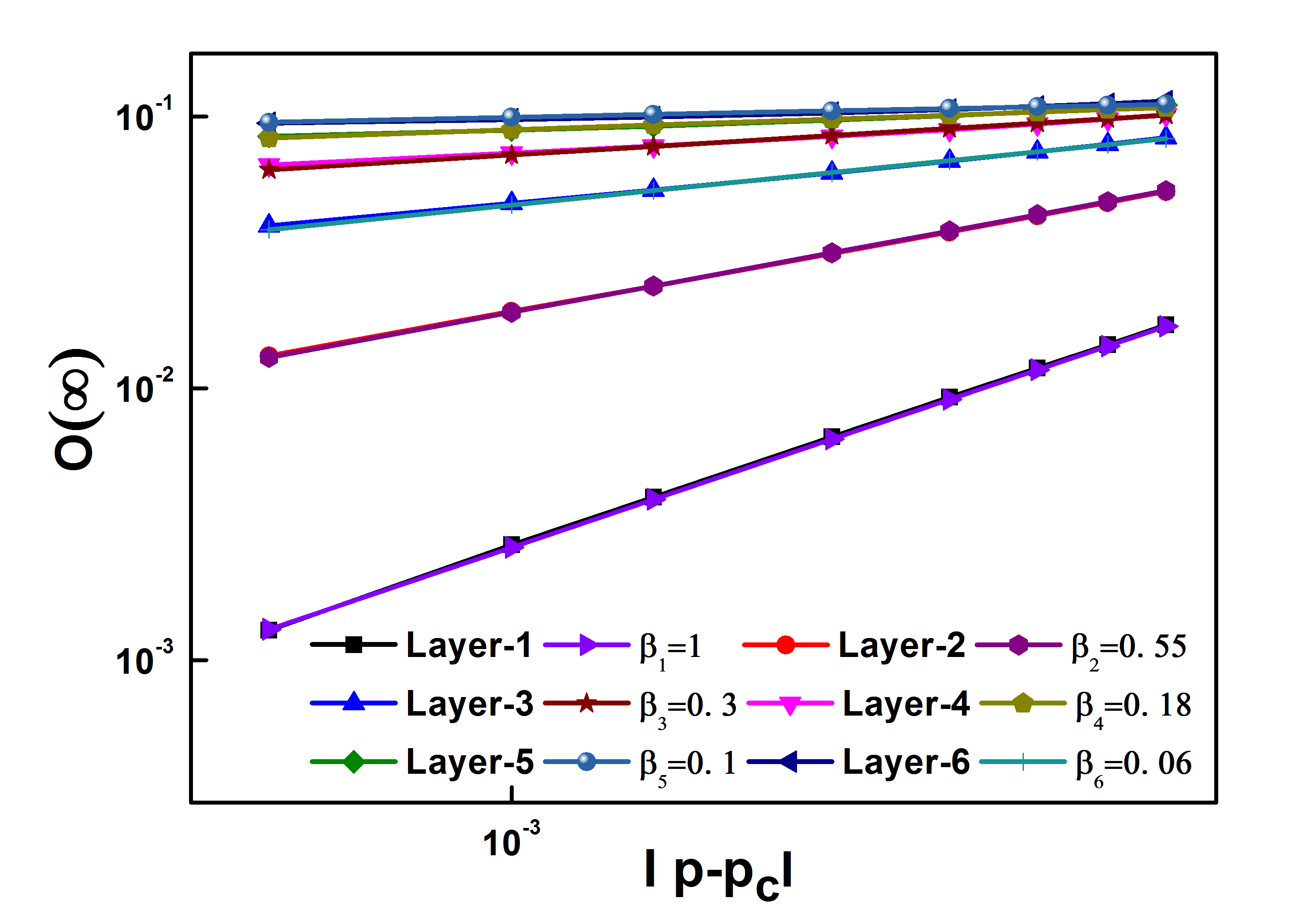}
}
\caption{The $O_l(\infty)$ is plotted for various values of
         $\Delta$ ranging from $0.0005-0.0065$
for various layers (from bottom to top). The behavior can be appoximated
as $O_l(\infty)\propto \Delta^{\beta_l}$ with
$\beta_l=\nu_{\parallel, l}\delta_l$,  and
$\nu_{\parallel,1}=1$
$\nu_{\parallel,2}=1.1$
$\nu_{\parallel,3}=1.2$
$\nu_{\parallel,4}=1.44$
$\nu_{\parallel,5}=1.6$
$\nu_{\parallel,6}=1.92$.}
\label{Fig:4}
\end{figure}

To understand this behavior, we write mean-field equations for
various layers.
The mean field equation for directed percolateion is given by Eq. 3.6
in \cite{henkel2008non}.
$\partial_t\rho_1(t)=\tau\rho_1(t)-g\rho_1(t)^2$.
for the critical point $\tau=0$, $\rho_1(t)={\frac{1}{c+gt}}$ where
$c=(\rho_1(0))^{-1}$. Thus $\delta_1=1$.
For $\tau>0$ $\rho_1(t)\sim {\frac{\tau}{g}}$
as $t\rightarrow \infty$ implying $\beta_1=1$ and hence $\nu_{\parallel}=1$
as expected in mean field limit.
We heuristically write equations for different layers as
\begin{align}
\partial_t\rho_1(t)=\tau\rho_1(t)-g\rho_1(t)^2 \nonumber \\
\partial_t\rho_2(t)=\tau\rho_2(t)-g\rho_2(t)^2+\rho_1(t)\nonumber\\
\vdots\nonumber\\
\partial_t\rho_l(t)=\tau\rho_l(t)-g\rho_l(t)^2+\rho_{l-1}(t)\nonumber\\
\vdots\nonumber\\
\partial_t\rho_L(t)=\tau\rho_L(t)-g\rho_L(t)^2+\rho_{L-1}(t)
\label{eq:}
\end{align}

We simulate these equations at the critical point $\tau=0$ using
fourth-order Runge-Kutta method with $h=0.01$ with 
$\rho_i(0)=0.9$ for $1\le i\le L$.
Asymptotically, we observe a power-law decay of
order parameter as $\rho_l(t)\sim t^\delta_l$ with $\delta_l=2^{1-l}$.
These plots are shown in Fig. 5. Thus the hierarchy of mean-field equations
explains the order density decay exponent at $p=p_c$ very well.

\begin{figure}[hbt!]
\scalebox{0.3}{
\includegraphics{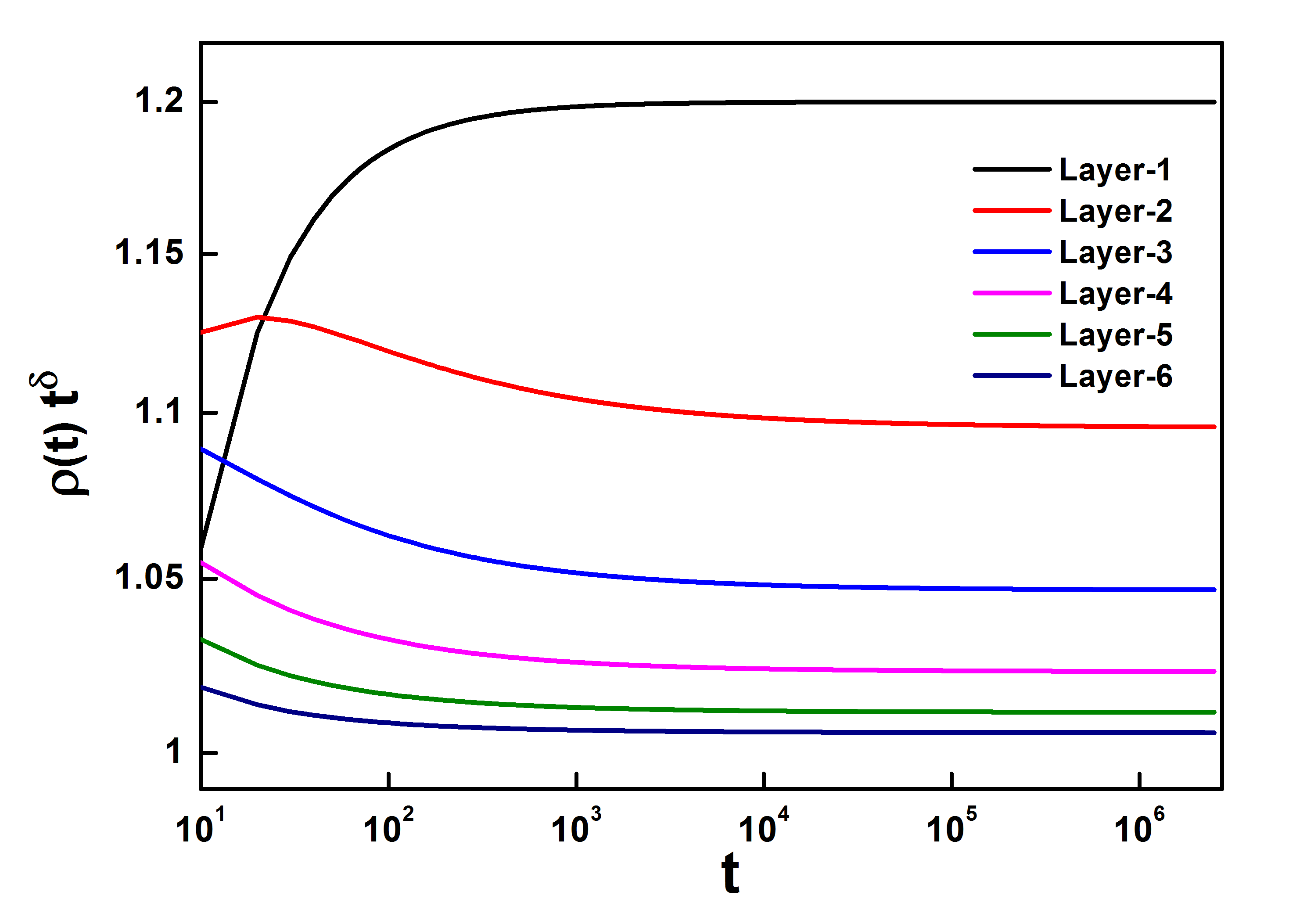}
}
\caption{The $t^{\delta_l}\rho_l(t)$ is plotted as function of time $t$
 for various layers (from top to bottom).}
\label{Fig:5}
\end{figure}

However for $\tau>0$, the behaviour does not match with random multiplex
described above. In an analogous manner, we propose $\rho_l(\infty)\propto \tau^{\beta_l}$ and   obtain $\beta_l=\delta_l$. 
Thus  $\nu_{\parallel}=1$ for
all layers which are expected for the mean-field system. This is not reproduced
for random network multiplex. The reason may be long crossover times or
the mean-field limit may be approached for very large values of $k$.
We have noted above that it is not necessary that non-equilibrium
systems on random networks show a transition in the mean-field class.\\

\begin{center}
{\textbf{B.  1-dimensional network}}
\end{center}
We also consider the case in which each layer has internal
connections like a d-dimensional cartesian lattice. Let us
consider the case of 1-d lattice and $L$ layers.
We study the system for $L=4$. We carry out the simulations
for $N=2\times 10^5$ and averaged over $220$ configuration.
The critical point $p_{c}=0.70548515$ is known\cite{henkel2008non} and is
the same for all layers of network.
As expected, there is clear power-law decay of order
parameter with critical exponent $\delta=0.159$ for the
first layer. (see fig.6)
This behavior is expected.
This absorbing phase transition 
is the same as in the DP class of 1-D lattice for
the top layer.
\begin{figure}[hbt!]
\scalebox{0.32}{
\includegraphics{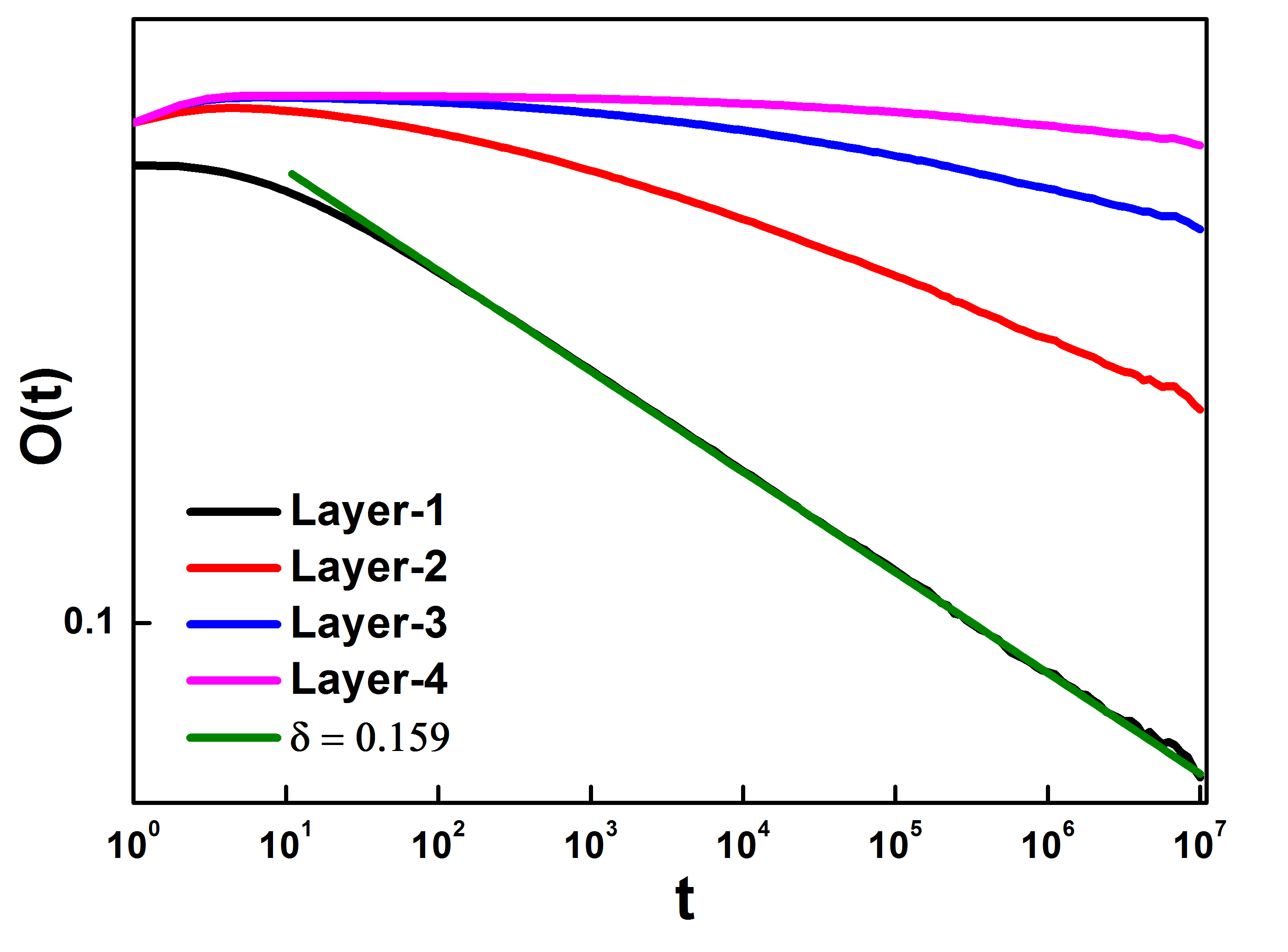}
}
\caption{We plot $O_l(t)$ as function of time $t$ 
for 1-D network (from bottom to top) for 
 $N= 2 \times 10^{5}$ and
$p=P_{c}= 0.70548515$. The order
 parameter decay exponent for top layer is $\delta=0.159$.}
\label{Fig:6}
\end{figure}
However, the decay of order parameter for layers below
the top layer is not a power-law decay. It is better fitted by a stretched
exponential. Except first layer, all other layers show a  stretched exponential decay
of order parameter as $\rho_l(t)\propto  \exp(-B_lx^{c_l})$
and the value of $c_l$ increases with $l$ (see fig.7). The values of $C_{l}$ 
are $ 0.09 ,0.16 $ and $ 0.24 $ within $3\%$ 
for the second, third, and fourth layers. This behavior is
confirmed by fitting using standard software such as
Origin\cite{edwards2002origin} and using a fit function in Gnuplot
which uses an implementation of the nonlinear least-squares(NLLS) 
Marquardt-Levenberg
algorithm\cite{ranganathan2004levenberg}.
\begin{figure}[hbt!]
\scalebox{0.16}{
\includegraphics{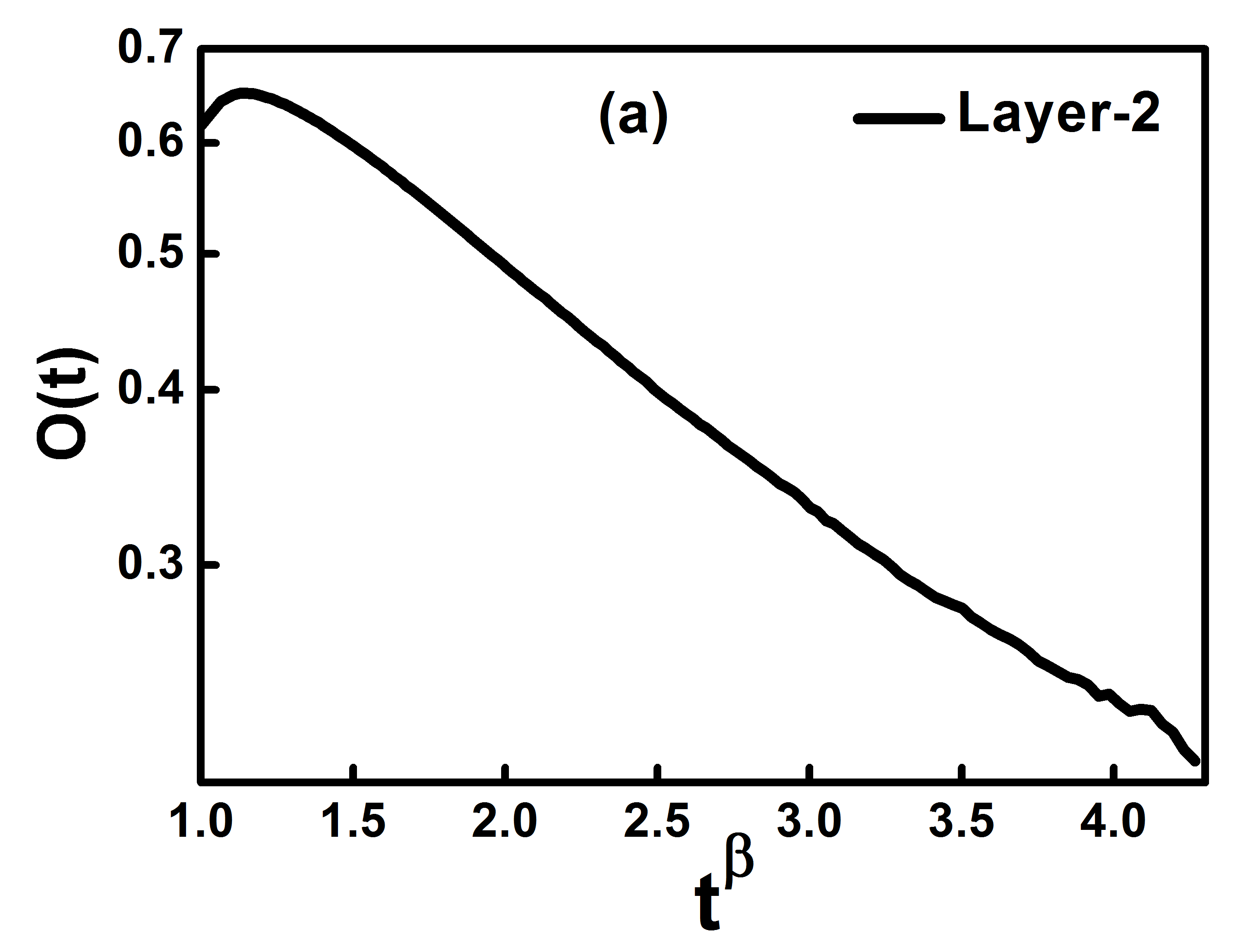}
}
\scalebox{0.16}{
\includegraphics{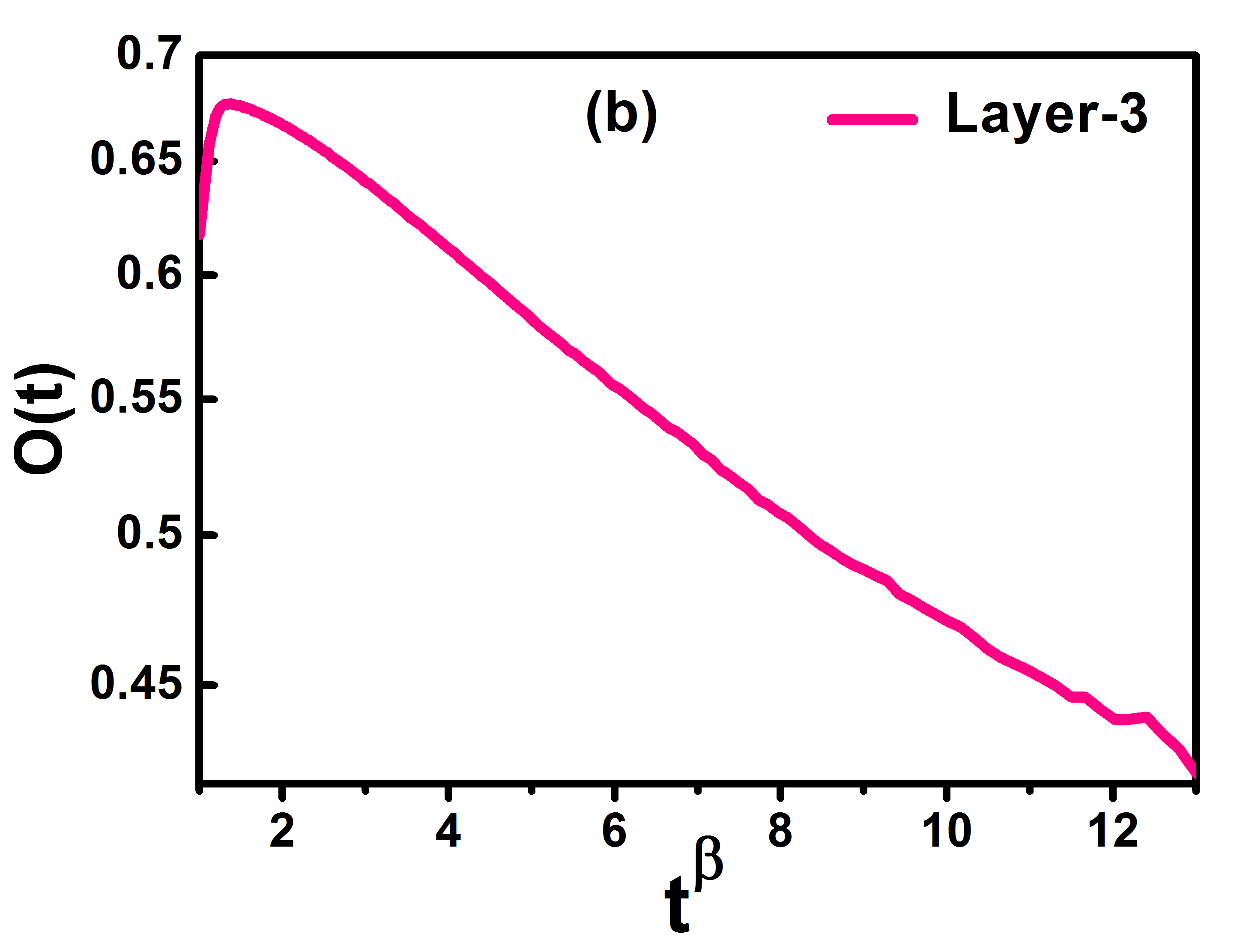}
}
\scalebox{0.16}{
\includegraphics{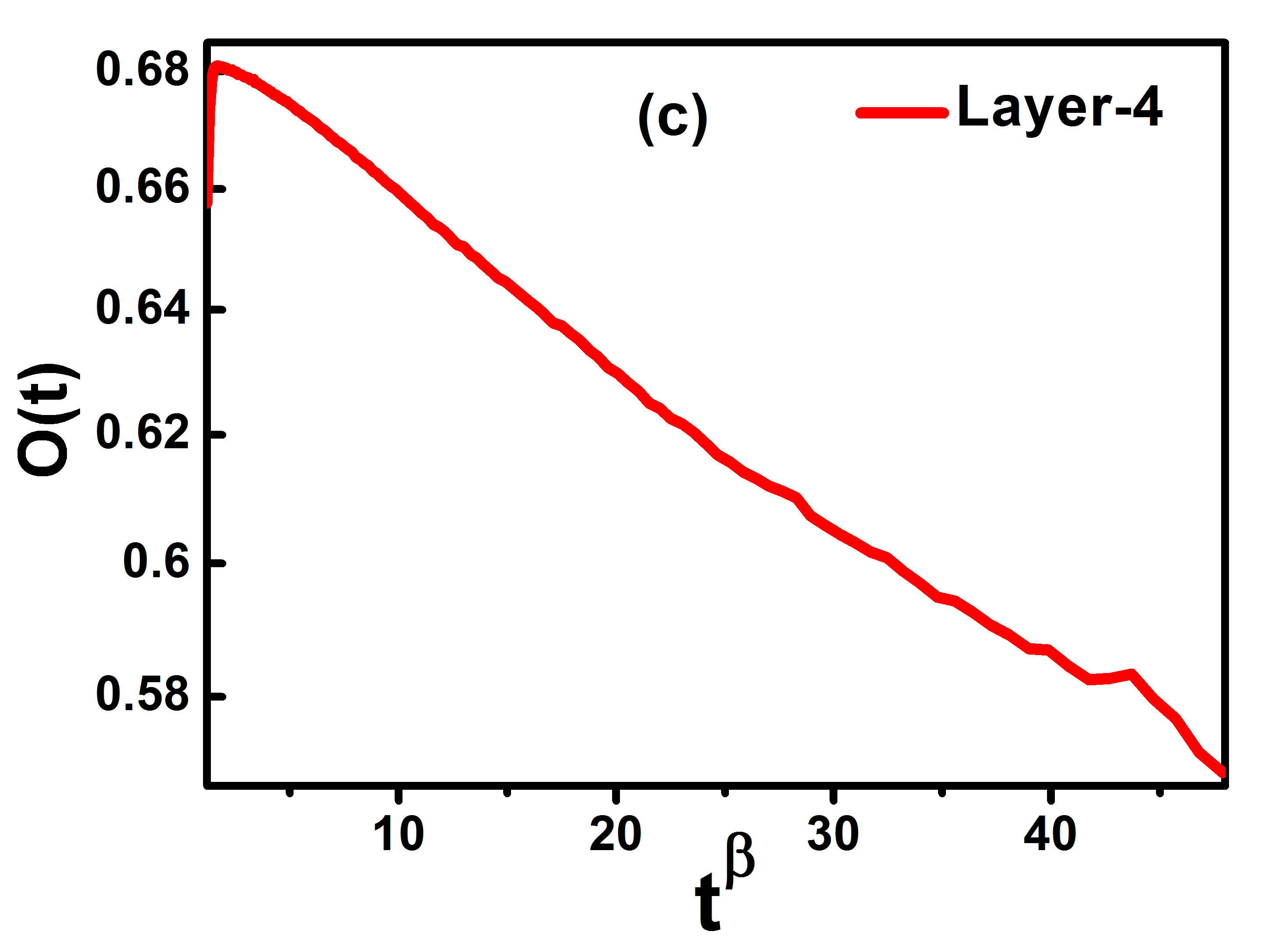}
}
\caption{For 1D, we plot $O_{j}(t)$ vs. $t^{\beta}$ on
semi-logarithmic scale for $j\ne 1$ at $p=p_c$.
A clear straight line shows that decay is well described by 
stretched exponential.
(a)  j=2, $\beta=0.09$  (b)  j=3, $\beta=0.16$ 
(c)  j=4, $\beta=0.24$ }
\label{Fig:7}
\end{figure} \\

\begin{center}
{\textbf{C.  2-dimensional network}}
\end{center}
We carry out similar investigations for the case where the
network in a given layer is 2-d.
We simulate
$N\times N$ lattice in a given layer with $N=2.5\times 10^3$
at $p=p_c=0.34457$.
We averaged over $85$ configuration and consider 4 layers.
The order parameter show a power-law decay with exponent $\delta= 0.45$
for the first layer as expected for  DP class(see fig.8).
However, others layers bend downwards on a 
log-log scale and decay is faster than power law.
As in the case of 1-D, it is described by
a clear stretched exponential decay. 
For 2D network the values of $C_{l}$ are $ 0.12,
0.23 $ and $ 0.41 $ within $1\%$
for second, third, and fourth layers. The plots
are shown in figure 9.
For $l=4$ in 2-D, curvature indicates the 
the possible presence of strong nonlinear
corrections to stretched exponential fit.
We note that stretched exponential is a very poor fit 
$l\ne1$ for random network.

\begin{figure}[hbt!]
\scalebox{0.32}{
\includegraphics{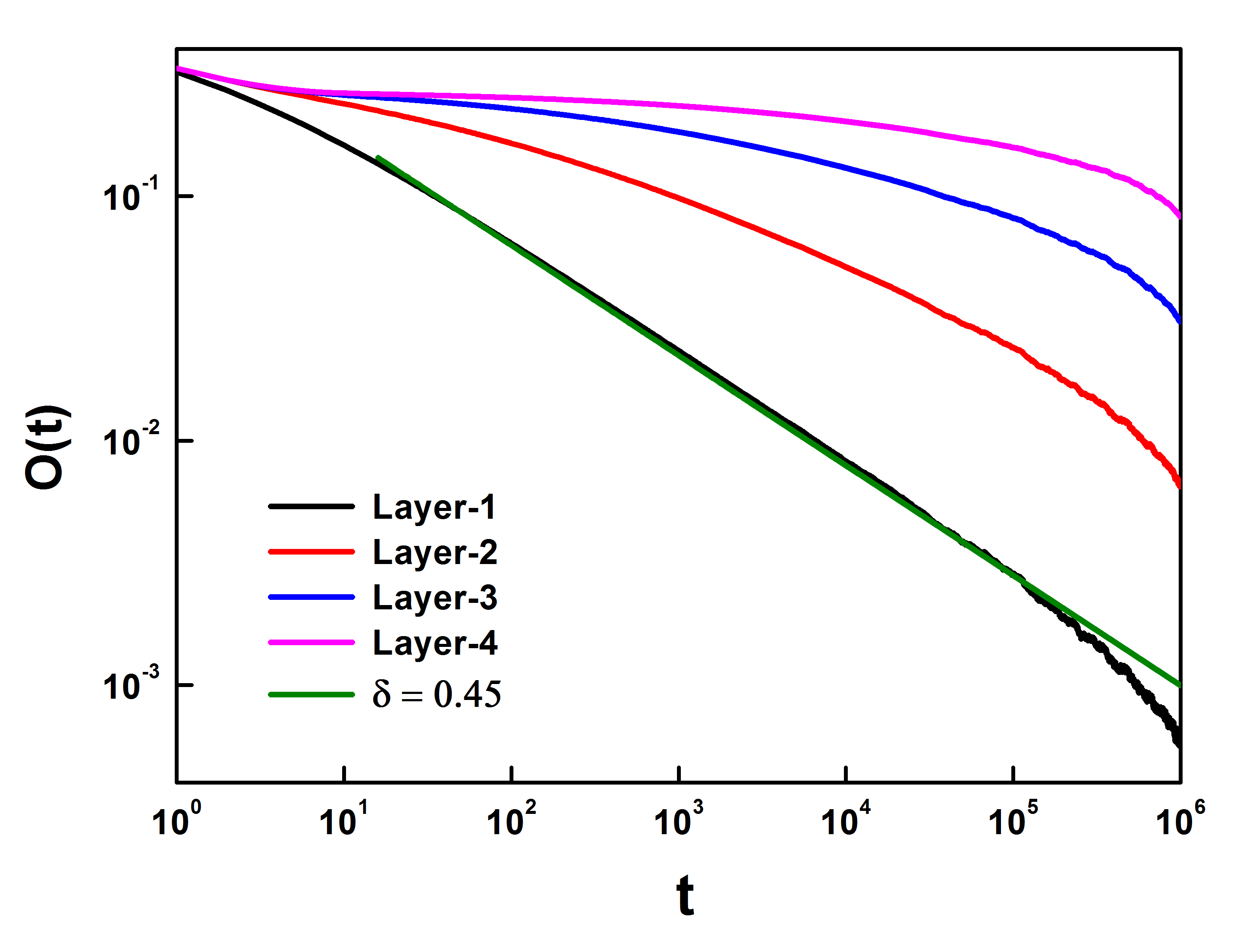}
}
\caption{We plot $O_j(t)$ as function of time $t$
for 2-D network (from bottom to top) of size $N= 2.5 \times 10^{3}$
at $p=P_{c}= 0.34457$. The decay exponent for first layer is $\delta = 0.45$.}
\label{Fig:8}
\end{figure}
\begin{figure}[hbt!]
\scalebox{0.16}{
\includegraphics{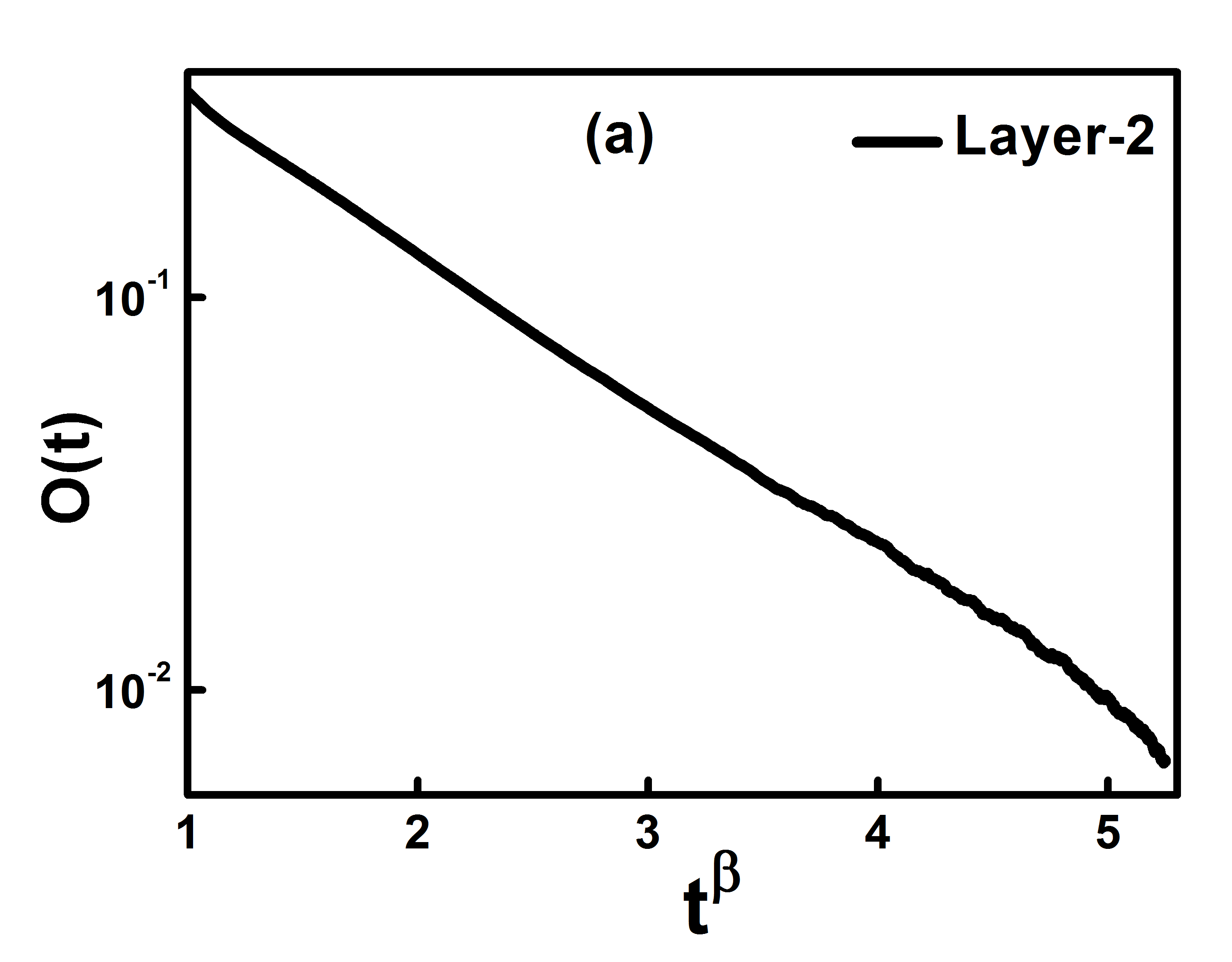}
}
\scalebox{0.16}{
\includegraphics{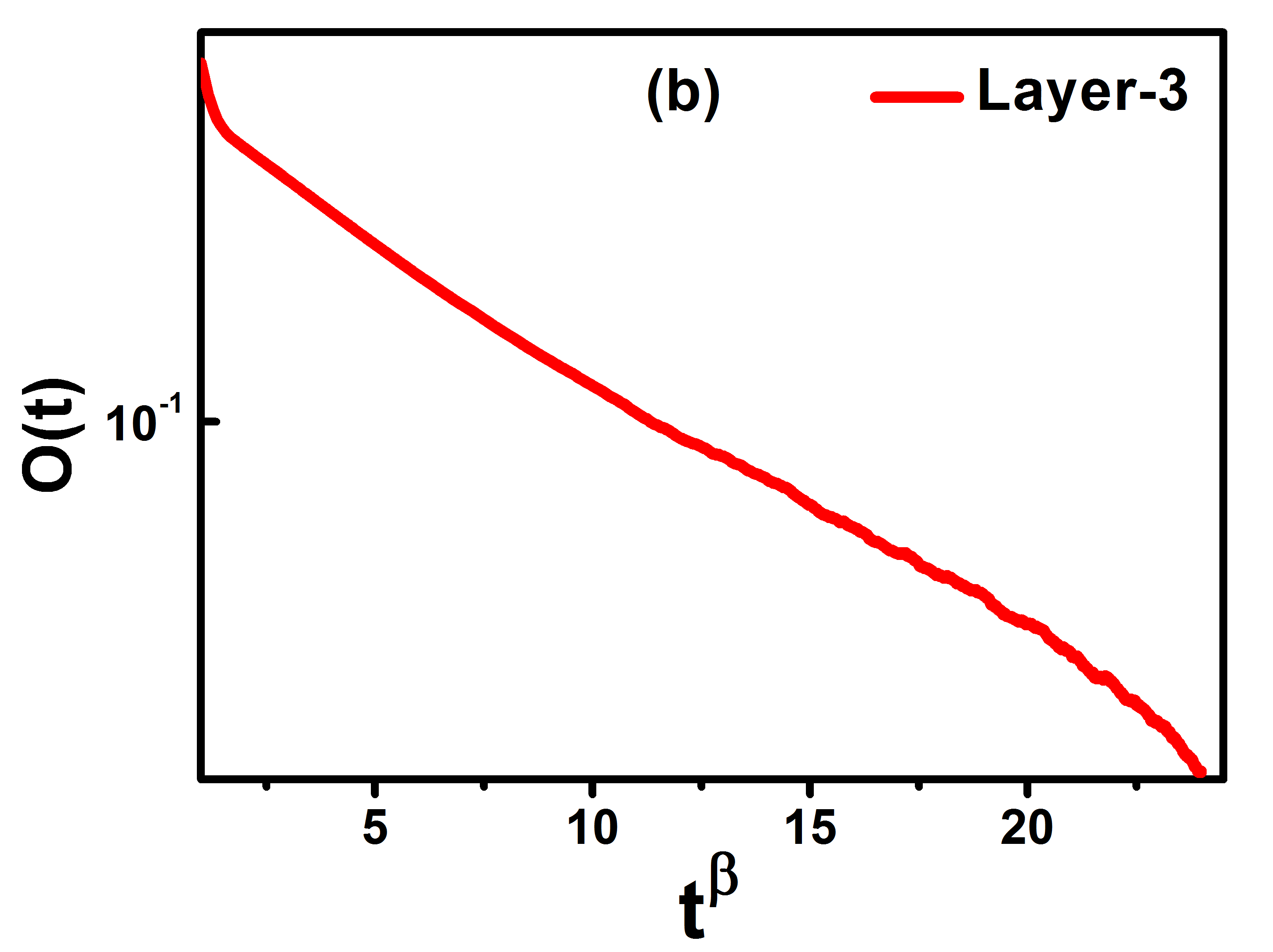}
}
\scalebox{0.16}{
\includegraphics{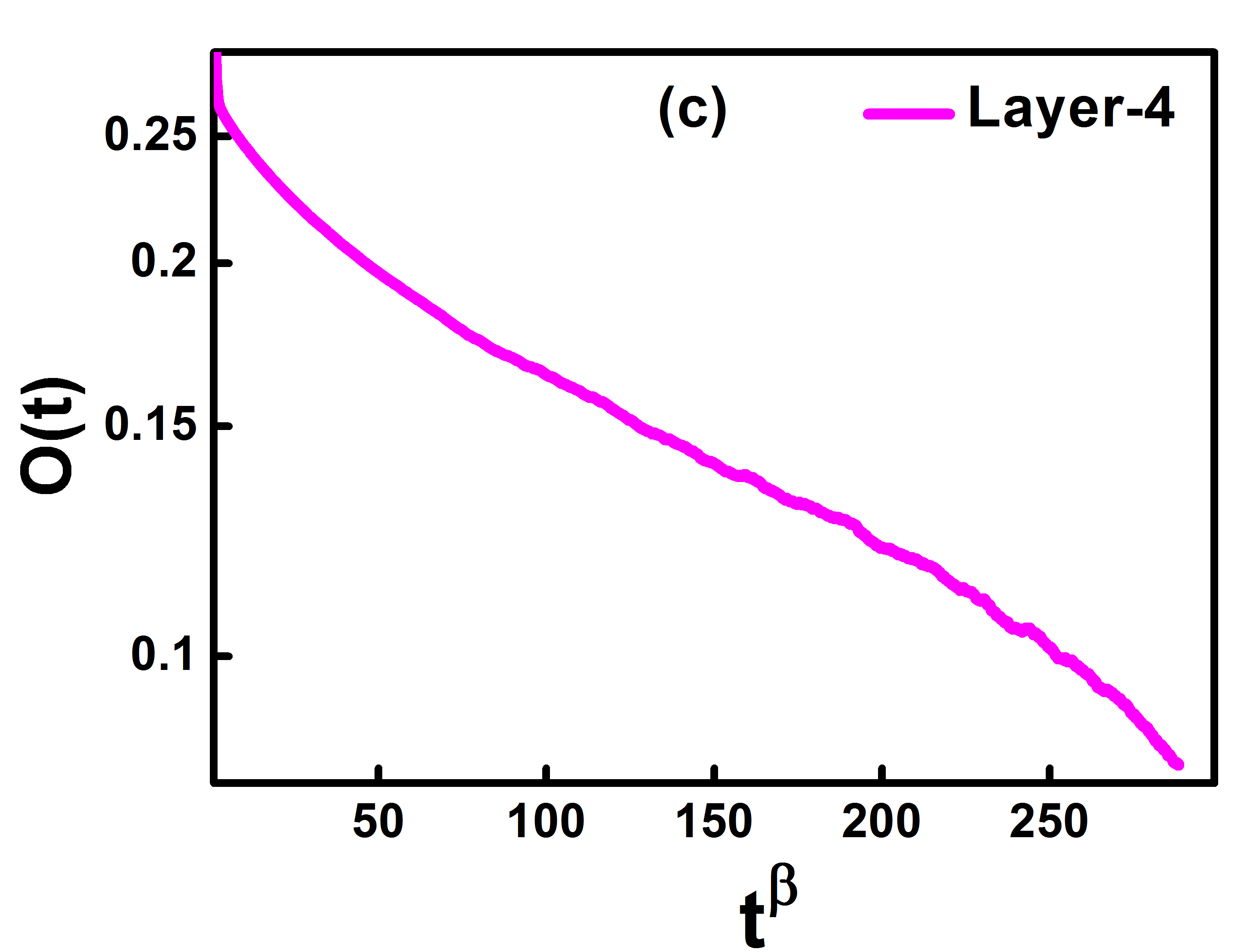}
}
\caption{We plot $O_{j}(t)$ vs. $t^{\beta}$ on
semi-log scale for $j\ne 1$ at $p=p_c$. Data is well fitted by
stretched exponential 
(a) j=2, $\beta=0.12$ (b) j=3, $\beta=0.23$ (c)  j=4, $\beta=0.41$ }
\label{Fig:9}
\end{figure}
\begin{center}
{\textbf{III.  Summary}}
\end{center}
In this paper, we discussed three systems i.e. random system, 1-D, and
2-D system. In these systems, each layer is connected to the layer above it in a unidirectional manner. The top layer
has no connection to any other layer.
The contact process in this system is defined in the following manner. Any
site becomes active with probability $p$ if any of the connected sites
is active.
The critical point for the top layer is
well known and the critical point is expected to be the same for entire
network.
We compute the fraction of active sites
$O_l(t)$ in a
given layer $l$ as an order parameter.  

(a) In a random network,
we find that there is a power-law decay of order parameter at each layer
for $p=p_c$
and the decay exponent is half of the previous layer. Since a well-defined
order parameter decay exponent is observed, we compute other exponents
such as finite-size scaling and off-critical scaling. We find that
the dynamic exponent $z=0.5$ for all layers is not the mean-field exponent. 
The saturation value
of order parameter for various layers  scales as $\Delta^{\beta_l}$
where $\beta_l=\delta_l\nu_{\parallel,l}$. Even the value of $\nu_{\parallel,l}
\ne 1$ except the first layer which is a departure from mean-field. We propose
a system of hierarchy of differential equations that correctly
reproduces the behavior at a critical point for all layers, but not the 
behavior in fluctuating phase.

(b) In 1-D and 2-D networks, the absorbing phase transition
in the first layer leads to a power-law decay of order parameter
only in the top layer. We find that other layers show a stretched exponential decay.
As expected, the power-law decay exponent of the first
layer is the same as to DP in 1-D or 2-D lattice. 
However, the decay is not described by power law for other layers.
It is better fitted by the stretched exponential.\\
\begin{center}
{\textbf{ACKNOWLEDGMENTS}}
\end{center}
PMG thanks DST-SERB (CRG/2020/003993) for financial assistance.
MCW thanks the Council of Scientific and Industrial Research 
(C.S.I.R.), SRF (09/128(0097)/2019-EMR-I).

\bibliographystyle{apsrev4-1.bst}
\bibliography{ref.bib}

\end{document}